\documentclass{PoS}

\title{A Cosmic-ray Electron Spectrum with VERITAS}

\ShortTitle{A Cosmic-ray Electron Spectrum with VERITAS}

\author{\speaker{David Staszak} for the VERITAS Collaboration\thanks{veritas.sao.arizona.edu}\\
        McGill University\\
        E-mail: \email{staszak@physics.mcgill.ca}}

\abstract{
Cosmic-ray electrons and positrons (CREs) at GeV-TeV energies are a unique probe of our local Galactic neighbourhood.
CREs lose energy rapidly via inverse Compton scattering and synchrotron processes while propagating in the Galaxy, effectively placing a maximal propagation distance for TeV electrons of order $\sim$1 kpc.
Within this window, detected CREs can come from only a handful of known, nearby astrophysical sources capable of exciting CREs to that energy or from more exotic production mechanisms, like particle dark matter.
HESS, and later MAGIC, have shown that ground-based imaging atmospheric Cherenkov telescopes have the capability to measure CREs into the TeV band.
In this proceedings we'll discuss the status of a VERITAS measurement of the electron plus positron cosmic ray spectrum.
}

\FullConference{The 34th International Cosmic Ray Conference,\\
		30 July- 6 August, 2015\\
		The Hague, The Netherlands}

\begin{document}

\section{Introduction}

Despite contributing only a small fraction of the overall cosmic-ray flux, cosmic-ray electrons and positrons (CREs) provide an important and unique probe of our local Galactic neighborhood.
CREs lose energy rapidly via inverse Compton scattering and synchrotron processes while propagating in the Galaxy, effectively placing a maximal propagation distance for TeV electrons of order $\sim$1 kpc\cite{KOBA}.
CREs at TeV energies provide a direct measurement of the local cosmic-ray accelerators and diffusion.

The Fermi-$LAT$\cite{LAT}, and more recently AMS\cite{AMS}, have measured the CRE spectrum with high statistics using satellite-based experiments up to energies of several hundred GeV.
However, above those energies these instruments run into difficulty from the combination of the steep CRE spectrum and their relatively small acceptance.
Ground-based experiments that utilize the Imaging Atmospheric Cherenkov Technique (IACT) have the capability to extend the CRE spectrum to higher energies due to their large 
collection area ($\sim10^{5}$m$^{2}$).
HESS\cite{HESS1}\cite{HESS2} and MAGIC\cite{MAGIC} have demonstrated this ability and previously measured the CRE spectrum from the ground up to TeV energies.
Their results provide evidence of at least one nearby CRE source and agree with satellite measurements within systematical uncertainties where there is energy overlap.
The combined picture that has emerged is one where the CRE spectrum is mostly flat and can be described by a simple powerlaw from $\sim$10 GeV up to just below $\sim$1 TeV, above which 
HESS has measured a spectral steepening.
MAGIC data\cite{MAGIC} are consistent with a single power-law up to $\sim$3 TeV.

Additionally, the positron fraction spectrum, $\phi(e^{+})/(\phi(e^{-})+\phi(e^{+}))$, has now been measured above 10 GeV by the HEAT\cite{HEAT}, PAMELA\cite{PAM}, Fermi-$LAT$\cite{LATfrac}, and AMS\cite{AMSfrac} experiments.
This ratio is found to rise with increasing energy up to $\sim$200 GeV, above which it appears to flatten out.
Positrons are mainly produced in secondary interactions between cosmic rays and the interstellar gas.
These results point to the possible existence of an additional local source of positrons on top of secondary production.
Explanations could include additional production from nearby standard astrophysical objects, such as pulsars or supernova remnants, or more exotic production mechanisms, such as the annihilation or decay of particle Dark Matter.
Conversely, recent studies have also proposed a more detailed propagation model\cite{prop} or a better accounting of secondary production\cite{waxman} to describe the results.
A full understanding of this situation will require detailed input about both the positron fraction and the CRE spectrum.
While the excitement of the unexpected excess found in earlier ATIC data\cite{ATIC} is now largely over, a high statistics measurement of the CRE spectrum to TeV energies will help us build a clear picture of our local CRE emitters.

\section{Methods}

Data presented here were collected by the VERITAS telescope array located at the Fred Lawrence Whipple Observatory (FLWO) in southern Arizona (31$^{\circ}$ 40$'$N, 110$^{\circ}$ 57$'$W,  1.3km a.s.l.).
VERITAS is a ground-based array of four telescopes sensitive to gamma- and cosmic-rays above $\sim$100 GeV.
While VERITAS is primarily a gamma-ray instrument, CREs are a diffuse source across the sky and so are collected during all astrophysical observations. %and buried in the background for 

\begin{figure}[t]
\begin{center}
\includegraphics[width = 4.5in]{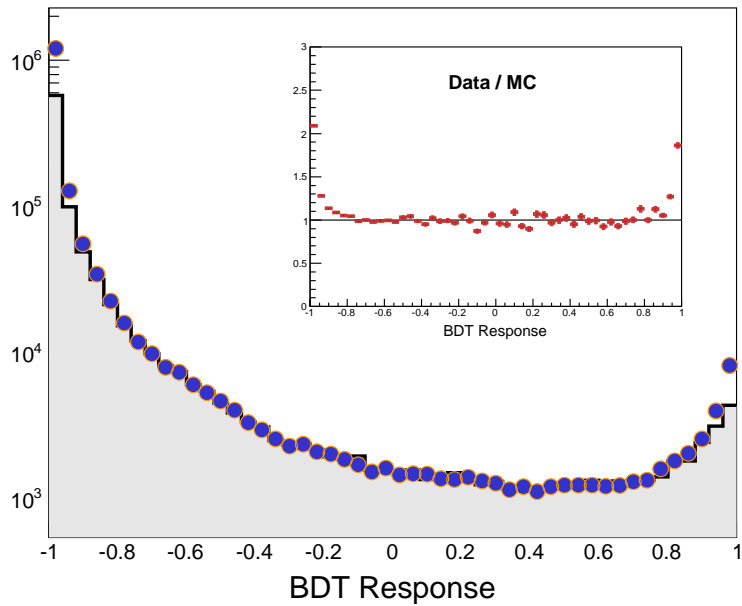}
\caption{BDT response for proton MC (gray filled area) and the full dataset (blue points) for the energy range 630 GeV to 1 TeV. 
The insert shows the ratio, data/MC, over the same range.
The agreement is very good except close to the limits of the distribution.
As we approach the side dominated by background-like events, -1.0, we find an excess in the data over the proton MC.
We expect an excess here from helium and higher-Z primaries, particularly since helium makes up $\sim20\%$ of the cosmic-ray flux.
Likewise, as we approach the side dominated by signal-like events, 1.0, we find an excess in the data over the proton MC that arises from the CREs measured in this study.}
\label{dataMC}
\end{center}
\end{figure}

\begin{figure}
\begin{center}
\includegraphics[width = 5.5in]{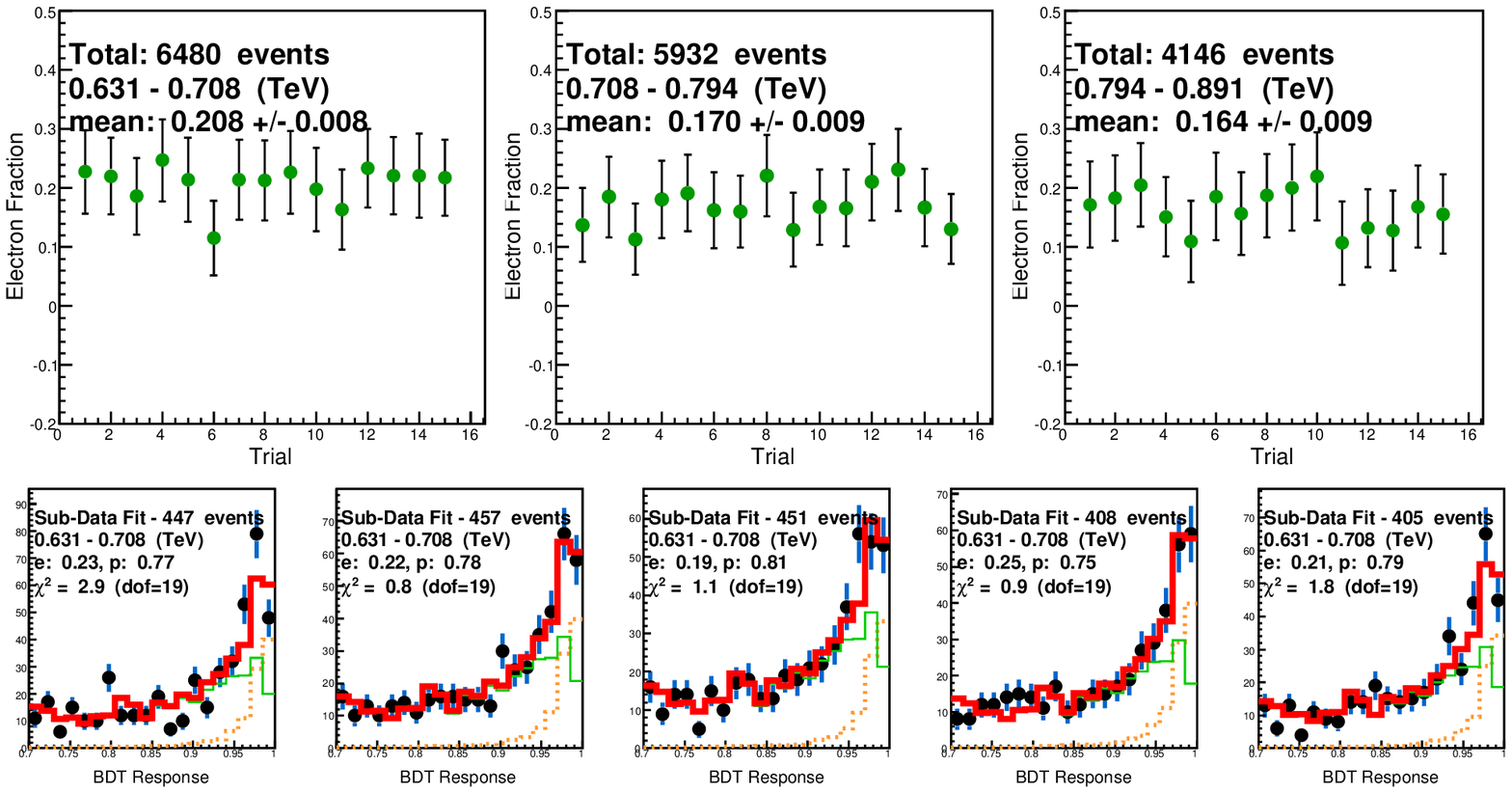}
\caption{The top three panels present the fit results for each of the 15 trials in three energy bins.
The x-axis represents the trial number and the y-axis represents the measured electron fraction (the individual trial error bars are the 1$\sigma$ errors from the fit).
Also shown is the mean value and uncertainty on the mean for each bin.
The bottom five panels show the individual fit results for the first five trials in the 631$-$708 GeV energy bin
(the black points are data, the best fit line is wide and red, the proton MC line is thin and green, and the electron MC line is dashed and orange).
For each fit we list the best fit electron and proton fractions and the $\chi^{2}/dof$ for the comparison between the data and best fit.
}
\label{METHOD}
\end{center}
\end{figure}

To isolate this signal, we apply strict analysis cuts to select only the best reconstructed events under pristine weather conditions.
Data events must have four good telescope images and reconstruct within the inner $1^{\circ}$ of the camera (the VERITAS field of view is $3.5^{\circ}$).
A strict cut is placed on the distance between the array center and the reconstructed array core, $coreR<200$ m.
Only extragalactic fields are considered in order to reduce the contamination from the Galactic Plane diffuse gamma-ray flux. 
Additionally, all detected or candidate gamma-ray sources within each field of view are excluded. 
To reduce the number of detector configurations in our simulations, we selected only data collected between September, 2009 and July, 2012, which are the dates of the two major VERITAS hardware upgrades (telescope relocation and PMT replacement, respectively).
We also restrict the mean zenith angle of the data to be between $65^{\circ}  - 75^{\circ}$, where we exclude data runs with mean values outside of this range (a data run is typically $\sim$20 minutes).
To accumulate sufficient Monte Carlo statistics, we generated simulations at a single zenith angle, $70^{\circ}$. 
This restricted data zenith range ensures the level of data/MC agreement necessary in this analysis.
296 hours of live-time remain after all these cuts.

We rely heavily on our Monte Carlo for interpretation and signal extraction.
To simulate the electromagnetic and hadronic showers, we used Corsika 6.970\cite{corsika}, with the QGSJetII.3 and URQMD 1.3cr underlying event generators, and GrISUDet 5.0.0\cite{grisu} for the VERITAS detector response.
We generated electrons, protons, and helium showers with a $4^{\circ}$ radius on the sky to approximate the isotropic and diffuse cosmic-ray flux.
Larger simulation radii were tested and found to not improve this approximation.

For signal and background discrimination we use Boosted Decision Trees (BDTs) that were integrated into the standard VERITAS analysis chain using the ROOT TMVA\cite{TMVA} framework.
The BDTs were trained with a diffuse electron MC sample (signal) and a representative sub-sample of the data chosen randomly from the full dataset (background).
We used four array-level shower variables in the training and event discrimination: MSCW, MSCL, $\chi^{2}(E)$, and the emission height.
MSCW and MSCL are variables based on the comparison of the spread along the major and minor axes of an ellipse fit to the camera images (length and width respectively) compared to expected values from simulations.
$\chi^{2}(E)$ represents the variability of the energy measurements in each of the four telescopes.
The emission height is the reconstructed height of the peak width of the shower.
We ensured that the trees were not overtrained by using a fraction of the full training signal and background samples to test the response.  
No overtraining was found and the trees and method were also used to successfully reconstruct the Crab Nebula energy spectrum as a cross-check.

Each data event is assigned a BDT response value, from $-$1.0 to 1.0, where higher values indicate that the event is more signal-like.
Figure \ref{dataMC} shows a comparison of the BDT response for the full dataset and proton MC.
The agreement is very good except near the limits of the distribution.
As we approach the side dominated by background-like events, $-$1.0, we find an excess in the data over the proton MC.
We expect an excess here from helium and higher-Z primaries, particularly since helium makes up $\sim20\%$ of the overall cosmic-ray flux\cite{pdg}.
We investigated the BDT response of helium MC events and they are found to peak at $-$1.0 and fall off faster than proton MC, in agreement with this interpretation.
As we approach the side dominated by signal-like events, 1.0, we find an excess in the data over the proton MC that arises from the CREs measured in this study.

We select only those events with BDT response values $>0.7$, which focuses on the region that contains the majority of the signal-like events and rejects the majority of the background-like events.
We then employ an extended likelihood fitting method to the BDT response within this region to extract the contribution of electron and proton events to the total.
This fit floats the electron and proton MC shapes relative to each other to find the best combined fit to the data.
Helium and higher-Z shower events are found to be sufficiently rejected by the BDT cut to ignore at first order.
To estimate the final electron fraction and its uncertainty, we divide the data into sub-samples and run the experiment several times (see Fig. \ref{METHOD}).
The mean of these separate trials is the final electron fraction with the uncertainty defined as the error on this mean.
This method removes some systematic biases from the fitting method and lets the data itself drive the statistical precision of the measured value.
One caveat of this technique is that fits in trials/bins with less than $\sim$100 data events do not return sensible results.
As a result we use fewer trials for higher energy bins where there are less statistics.
The highest bin presented here is the result of a single fit to 200 events; the uncertainty quoted is the 1$\sigma$ error from the fit.

\begin{figure}[t]
\begin{center}
\includegraphics[width = 5.0in]{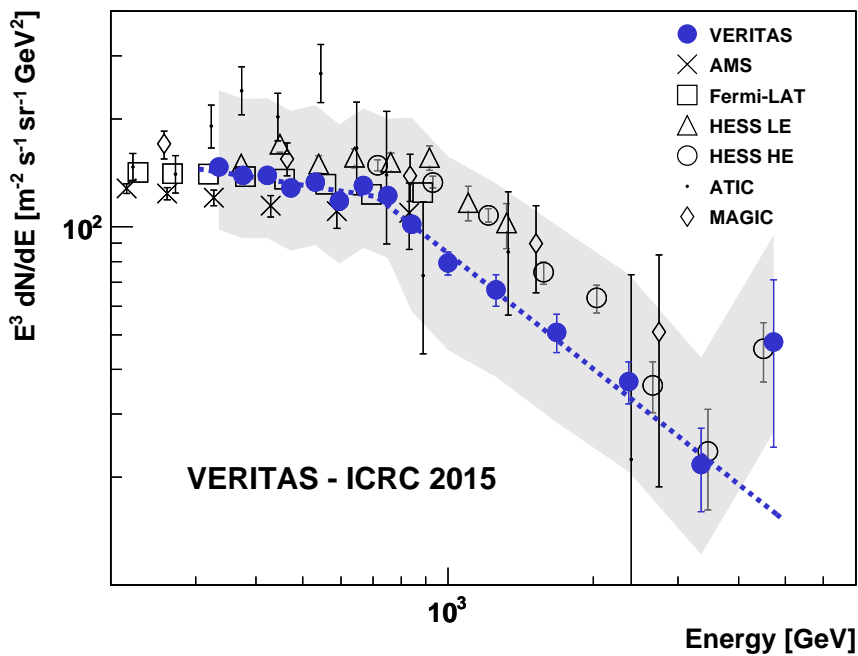}
\caption{VERITAS preliminary cosmic-ray electron spectrum as a function of energy (GeV) in solid blue circles.
The best fit to the data is represented as a dashed line and is found to be two power-laws with a break energy 710 $\pm$ 40 GeV and spectral indices of $-$3.1 $\pm$ 0.1$_{stat}$ ($-$4.1 $\pm$ 0.1$_{stat}$) below (above) the break.
The $\chi^{2}/$dof  of the fit is 0.9.
Shown for comparison are data from other experiments in the same energy range.  
The gray band represents the systematical uncertainty on the VERITAS measurement.}
\label{MONEY}
\end{center}
\end{figure}

\section{Results}

We show in Fig. \ref{MONEY} the preliminary VERITAS CRE energy spectrum spanning $\sim$300 GeV to $\sim$5 TeV.
The spectrum steepens at higher energies and is best described by two power-laws with a cutoff.
The best fit for this cutoff energy is found to be 710$\pm$40$_{stat}$ GeV, with best fit spectral indices below (above) this energy of $-$3.2 $\pm$ 0.1$_{stat}$ ($-$4.1 $\pm$ 0.1$_{stat}$).
The $\chi^{2}/dof $ of this fit is 9.71/11.
The gray band represents the systematical uncertainty, which is dominated by the $\sim20\%$ uncertainty on the VERITAS absolute energy scale.
This translates into a +64\%/$-$33\% (+98\%/$-$43\%) systematical uncertainty for a spectral index of $-$3.2 ($-$4.1).
We measure an additional $10\%$ systematical uncertainty above $\sim$1 TeV that quantifies hardware uncertainties at those energies.
This additional uncertainty is added in quadrature.

Due to the similarity of gamma and electron electromagnetic showers, we cannot rule out a significant contamination of gamma-ray events within our electron data.
However, Fermi-$LAT$ has now measured the diffuse extragalactic gamma-ray flux up to $\sim$800 GeV and finds it orders of magnitude below their own CRE measurement\cite{diffuseLAT}.
They additionally find evidence for a cutoff in the diffuse gamma-ray spectrum above a couple hundred GeV.

CRE results shown here qualitatively agree with prior ground-based and satellite-based measurements at similar energies.
Of the many experiments studying CREs, this result represents the second high-statistics measurement of a cutoff in the CRE spectrum around $\sim$1 TeV.
The precise measurement of this energy cutoff is an important parameter in any successful model our local CRE environment.
VERITAS has significantly more data on disk than what has been used in this study and work continues with the goal of extending our measurement out to even higher energies.
The CRE spectrum between 5 and 10 TeV is unexplored and this is the energy range where we expect to see spectral features from individual nearby astrophysical sources (if they are the dominant source).
We urge caution in over-interpretation of the uptick in the final VERITAS data point since this is within 2$\sigma$ of the best fit line.

\section*{Acknowledgements}

This research is supported by grants from the U.S. Department of Energy Office of Science, the U.S. National Science Foundation and the Smithsonian Institution, and by NSERC in Canada. We acknowledge the excellent work of the technical support staff at the Fred Lawrence Whipple Observatory and at the collaborating institutions in the construction and operation of the instrument.
Computations were made on the supercomputer Guillimin from McGill Univesity, managed by Calcul Quebec and Compute Canada. The operation of this supercomputer is funded by the Canada Foundation for Innovation (CFI), NanoQuebec, RMGA and the Fonds de recherche du Quebec - Nature et technologies (FRQ-NT).

The VERITAS Collaboration is grateful to Trevor Weekes for his seminal contributions and leadership in the field of VHE gamma-ray astrophysics, which made this study possible.


\begin{thebibliography}{99}
\bibitem{KOBA} T. Kobayashi et al., {\it The Most Likely Sources of High-Energy Cosmic-Ray Electrons in Supernova Remnants.} Astrophys. J. 601, 340, 2004.
\bibitem{LAT} M. Ackermann, et al.  {\it Fermi LAT observations of cosmic-ray electrons from 7 GeV to 1 TeV.}  Phys. Rev. D82, 092004, 2010.
\bibitem{AMS} M. Aguilar, et al. {\it Precision Measurement of the (e$^{+}$+e$^{-}$) Flux in Primary Cosmic Rays from 0.5 GeV to 1 TeV with the Alpha Magnetic Spectrometer on the International Space Station.}  Phys. Rev. Lett. 113, 221102, 2014.
\bibitem{HESS1} F. Aharonian, et al. {\it Energy Spectrum of Cosmic-Ray Electrons at TeV Energies.} Phys. Rev. Lett. 101, 261, 2008.
\bibitem{HESS2} F. Aharonian, et al. {\it Probing the ATIC peak in the cosmic-ray electron spectrum with H.E.S.S.} Astron. Astrophys. 508, 561, 2009.
\bibitem{MAGIC} D. Borla Tridon, et al. {\it Measurement of the cosmic electron plus positron spectrum with the MAGIC telescopes.} Proc. of the 32nd ICRC, 2011, arXiv:1110.4008.
\bibitem{HEAT} J.J. Beatty, et al. {\it New Measurement of the Cosmic-Ray Positron Fraction from 5 to 15 GeV.} Phys. Rev. Lett. 93, 241102, 2004.
\bibitem{PAM} O. Adriani, et al. {\it Observation of an anomalous positron abundance in the cosmic radiation.} Nature 458, 607, 2009.
\bibitem{LATfrac} M. Ackermann, et al. {\it Separate Cosmic-Ray Electron and Positron Spectra with the Fermi Large Area Telescope.} Phys. Rev. Lett. 108, 011103, 2012.
\bibitem{AMSfrac} M. Aguilar, et al. {\it High Statistics Measurement of the Positron Fraction in Primary Cosmic Rays of 0.5-500 GeV with the Alpha Magnetic Spectrometer on the International Space Station.} Phys. Rev. Lett. 113, 121101, 2014.
\bibitem{ATIC} J. Chang, et al. {\it An anomalous positron abundance in cosmic rays with energies 1.5-100 GeV.} Nature, 458, 607, 2008.
\bibitem{prop}  D. Gaggero, et al. {\it Three-Dimensional Model of Cosmic-Ray Lepton Propagation Reproduces Data from the Alpha Magnetic Spectrometer on the International Space Station.}  Phys. Rev. Lett. 111, 021102, 2013.
\bibitem{waxman} K. Blum, et al. {\it AMS 02 results support the secondary origin of cosmic ray positrons.} Phys. Rev. Lett. 111, 211101, 2013.
\bibitem{TMVA} A. Hoecker, et al.  {\it TMVA - Toolkit for Multivariate Data Analysis.} PoS ACAT 040, 2007, arXiv:0703039.
\bibitem{pdg} K.A. Olive et al. {\it Review of Particle Physics.} Chin. Phys. C38, 090001 (2014)
\bibitem{corsika} https://www.ikp.kit.edu/corsika/
\bibitem{grisu} http://www.physics.utah.edu/gammaray/GrISU/
\bibitem{diffuseLAT} M. Ackermann, et al. {\it The spectrum of isotropic diffuse gamma-ray emission between 100 MeV and 820 GeV.}  ApJ 799, 86, 2015.
\end{thebibliography}
\end{document}